\documentstyle[epsfig]{elsart}

\begin{document}

\thispagestyle{empty}
\begin{large}
\textbf{DEUTSCHES ELEKTRONEN-SYNCHROTRON}\\
\end{large}

DESY 03-165

October 2003

\begin{eqnarray}
\nonumber &&\cr \nonumber && \cr \nonumber &&\cr
\end{eqnarray}

\begin{center}
\begin{Large}
\textbf{Misconceptions regarding the cancellation of self-forces
in the transverse equation of motion for an electron in a bunch}
\end{Large}
\begin{eqnarray}
\nonumber &&\cr \nonumber && \cr
\end{eqnarray}

\begin{large}
Gianluca Geloni
\end{large}

\textsl{\\Department of Applied Physics, Technische Universiteit
Eindhoven, \\P.O. Box 513, 5600MB Eindhoven, The Netherlands}
\begin{eqnarray}
\nonumber
\end{eqnarray}
\begin{large}
Evgeni Saldin and Evgeni Schneidmiller
\end{large}

\textsl{\\Deutsches Elektronen-Synchrotron DESY, \\Notkestrasse
85, 22607 Hamburg, Germany}
\begin{eqnarray}
\nonumber
\end{eqnarray}
\begin{large}
Mikhail Yurkov
\end{large}

\textsl{\\Particle Physics Laboratory (LSVE), Joint Institute for
Nuclear Research, \\141980 Dubna, Moscow Region, Russia}

\newpage

\end{center}

\begin{frontmatter}

\journal{}

\title{Misconceptions regarding the cancellation of self-forces in the transverse equation of motion for an electron in a bunch}

\author[tue]{G.~Geloni}
\author[DESY]{E. L.~Saldin}
\author[DESY]{E. A.~Schneidmiller}
\author[DUBNA]{M. V.~Yurkov}

\address[tue]{Department of Applied Physics, Technische Universiteit
Eindhoven, The Netherlands}

\address[DESY]{Deutsches Elektronen-Synchrotron (DESY), Hamburg,
Germany}

\address[DUBNA]{Joint Institute for Nuclear Research, Dubna, 141980
Moskow region, Russia}

\begin{abstract}

As a consequence of motions driven by external forces, self-fields
originate within an electron bunch, which are different from the
static case. In the case of magnetic external forces acting on an
ultrarelativistic beam, the longitudinal self-interactions are
responsible for CSR (Coherent Synchrotron Radiation)-related
phenomena, which have been studied extensively. On the other hand,
transverse self-interactions are present too. At the time being,
several existing theoretical analysis of transverse dynamics rely
on the so-called cancellation effect, which has been around for
more than ten years. In this paper we explain why in our view such
an effect is not of practical nor of theoretical importance.

\end{abstract}
\end{frontmatter}

\clearpage

\setcounter{page}{1}

\section{Introduction}

Electron bunches with, among other characteristics, very small
transverse emittance and high peak current are needed for the
operation of XFELs \cite{t,l}. The bunch length for XFEL
applications is of the order of 100 femtoseconds. This is achieved
using a two-step strategy: first generate beams with small
transverse emittance by means of a RF photocathode and, second,
apply longitudinal compression at high energy using a magnetic
compressor. Self-interactions may spoil the bunch characteristics,
so that, in the last few years, simulation codes have been
developed in order to solve the self-consistent problem of an
electron bunch moving under the action of external magnetic fields
and its own self-fields. These codes show that, in the case under
examination, the bunching process can be treated in the zeroth
order approximation by the single-particle dynamical theory: in
fact we cannot neglect the CSR-induced energy spread and other
wake-fields when we calculate the transverse emittance, but we can
neglect them when it comes to the characterization of the
longitudinal current distribution because their contribution to
the longitudinal dynamics, in the high-$\gamma$ limit, is small.
As actual calculations are left to simulations, the understanding
of the physics of self-interactions is left to analytical studies,
which are usually based on a perturbative approach: the particles
move, in the zeroth order, under the influence of external guiding
fields and the zeroth order motion is used to calculate the first
order perturbation to the trajectory due to the self-interactions.
Of course, the perturbative approach will, in the most general
case, give different answers with respect to the self-consistent
approach as self-interaction effects get more and more important.
Nevertheless the method behind the two approaches is the same: the
only difference is due to the fact that the computational power is
enough to break the particle trajectories in sufficiently small
parts so that the first order perturbation theory gives a
satisfactory description of the bunch evolution within a given
trajectory-slice. Moreover it should be noted that  perturbation
techniques are important from a practical viewpoint too. In fact,
when it comes to facility design, one has to find a range of
parameters such that the emittance dilution is small, i.e. that
the self-consistent approach can be effectively substituted, from
a practical viewpoint, by the perturbative one. Partial analytical
studies have been performed, in the last twenty years, on
transverse self-fields, and a particular effect, called
cancellation, has become on fashion in the last few years
\cite{RUI,BOHN,DERB2,DERUI}. Analytical results obtained in those
papers led part of the community to believe that such an effect is
very fundamental and of great practical importance. We disagree
with this viewpoint. In this paper we explain our reasons. In
Section 2 we trace a short, conceptual history of the cancellation
effect. Then we move to the description of the current state of
the theory of this effect in Section 3. After this, in Section 4,
we explain our arguments against the fact that the cancellation
effect is of any practical and theoretical interest. Finally, in
Section 5, we come to conclusions.

\section{Twenty years of history}

The history of the so called cancellation effect is a long-dated
one.  A short summing up can be of interest here since on the one
hand it gives the reader a more thorough overview of the issue,
while on the other it constitutes a self-contained example of how
scientific progress works in time, by trial and error and,
sometimes, misconception and misunderstanding.

The story begins about twenty years ago with \cite{TALM}. In that
pioneering work the transverse self-force, i.e. the Lorentz
self-force in the transverse direction lying on the orbital plane,
is calculated for a test particle in front of a beam of zero
transverse extent moving in a circle \footnote{We will refer to it
with the name "tail-head interaction", since electromagnetic
signals form the bunch tail interact with the head.}, showing a
singular result for $\theta_{\mathrm{min}} = (s_{\mathrm{min}}/R)
\rightarrow 0$, $s_{\mathrm{min}}$ being the minimal arc length
from the nearest retarded source point to the test particle and
$R$ being the radius of the circle (see Eq. (8) in \cite{TALM}).
The singularity has a logarithmic character arising from the
particular model selected, "is due to 'nearby' charges and is
removed as the beam is given a transverse size" (quoted from
\cite{TALM}). This force constituted, in the late 80s, a reason
for serious concern for the beam dynamics in electron storage
rings. In Eq. (42) of \cite{LEE}, a particle off axis of a
quantity $x$ with respect to the design orbit  feels a total
centrifugal Lorentz force with logarithmic dependence on $x$ from
an "unshielded ring of charge of vanishing height and thickness"
(cited from \cite{LEE}, i.e. the same beam with zero transverse
extent analyzed in \cite{TALM}) :

\begin{equation}
F = {2\lambda\over {R}} \left[\ln\left({{x\over{8R}}}\right) +
1\right] + O(x/R)~, \label{lee1}
\end{equation}

\noindent $\lambda$ being the constant bunch linear density.

It should be underlined that the logarithmic singularity in $x$
comes from the choice of a particular density distribution. Not
all distribution choices give singular behavior as $x \rightarrow
0$. For example, in \cite{TALM}, a singularity is first found as
$s_{\mathrm{min}} \rightarrow 0$; we found a similar result in
\cite{OURS1}. This is also linked to the particular choice of the
distribution, that is a line bunch. The singularity for $
s_{\mathrm{min}} \rightarrow 0$ has actually the same reason to be
as the one for $x\rightarrow 0$ when an horizontally off-axis
particle is considered. Simply, if we start from a line bunch
case, and we consider a test particle with $x=0$, then the
singular character of the field comes into play when
$s_{\mathrm{min}} \rightarrow 0$; when $x\ne 0$, one is allowed to
put $s_{\mathrm{min}} = 0$ and then a singularity is present in
the limit for $x\rightarrow 0$. However, this is due to the choice
of the density distribution and nothing else. There are many
choices (for example a gaussian 2D or 3D bunch) which do not give
singular fields. As a matter of fact, macroparticle simulations
are based on these kind of simplified distributions, because they
avoid the problem of singularities. From a practical viewpoint
then, when dealing with computational problem, one should remind
that macroparticle simulation do not have to cope necessarily with
singularities, which arise only in the case of particular particle
distribution choices: this fact should be kept in mind throughout
the reading of this paper.

As we read in \cite{LEE}, the transverse force was "the subject of
serious concern for its effect on dynamics in electron storage
ring", in that it spoils the cancellation between the electric
field term and the magnetic field, which one usually has in pure
space-charge problems for the ultrarelativistic regime: the
gradient of $F$ in the radial direction is, in fact, singular as
$\partial F/\partial r \sim \lambda/x$ \footnote{This is just
another way to say that the transverse force exhibit a logarithmic
singularity for $x\rightarrow 0$. }. In the same paper, it is
pointed out that this force is cancelled for long beams undergoing
a circular motion. In fact "a particle undergoing betatron
oscillations has simultaneous oscillation of its kinetic energy
[...] in curved geometry the kinetic-energy oscillation results in
a first-order dynamical term in the horizontal equation of motion,
which shifts the betatron frequency. For a highly relativistic
beam this additional term nearly cancels a term proportional to
the gradient of the CFSC" (quoted from \cite{LEE}), where the
CFSC, here, is just the Lorentz force: "we define the CSFC with an
overall minus sign to agree with the conventions of other authors:
$F=-(E_r+B_z)_b$" (quoted from \cite{LEE}), $r$ and $z$ referring
to a cylindrical coordinate system $(r, \theta, z)$ and the
subscript $b$ indicating "beam-induced components of fields".

\section{The "cancellation effect" today \label{sec:ruicanc}}

With the development of SASE FEL technology and the need of
magnetic chicanes for bunch compression, the scientific community
began to analyze this kind of effects in the case of short beams
moving in finite arc of circles as in the bends of magnetic
chicanes. The aim was to end up with a generalization of the
results in \cite{LEE} which was valid under more generic
assumptions: not only for coasting beams in storage rings, but
also for bunched beams in generic magnetic systems. A first step
in that direction was claimed in \cite{RUI} and in \cite{DERB2},
to end up with a final formulation \cite{DERUI}.

Let us briefly describe the arguments in \cite{RUI}. We track the
electron bunch in cylindrical coordinates with respect to the
center of a designed circular orbit of radius $R$ as in \cite{RUI}
or \cite{BOHN}: $\vec r = r \vec e_r+ r\theta \vec e_s$. Following
the Hamilton-Lagrange formulation in \cite{RUI} or \cite{BOHN} one
can write the transverse equation of motion for a test particle:

\begin{equation}
{d\over{dt}}\left(\gamma m {d r\over{dt}} \right) - \gamma_0 m c^2
\left({\beta_s^2\over{r}}-{\beta_s\beta_0\over{R}}\right) =
F_\bot(t) + {\beta_s^2 \Delta E(t)\over{r}}. \label{transverse}
\end{equation}

This is a well-known starting point for the study of the
transverse dynamical problem, and it is used in theoretical
studies as well as by codes like $\verb"TraFiC"^4$. In order to
solve the equation of motion one should specify initial conditions
for the particle positions and velocities at the magnet entrance,
for which there is freedom of choice, from a general viewpoint.
All treatments must use Eq. (\ref{transverse}) as a starting
point. Then, if the initial conditions are the same, the results
must be the same: our discussion is not centered on wether it is
allowed to make certain manipulations of Eq. (\ref{transverse})
using the cancellation scheme but simply on whether it is useful
or not to do so, both from a theoretical and computational
viewpoint.

The two terms on the right hand side of Eq. (\ref{transverse}) are
simply the Lorentz force in the transverse direction of motion,
and the deviation of the particle energy at time $t$, $\gamma(t) m
c^2$, from the design energy $\gamma_0 m c^2$. Note that the
latter term can be also expressed as

\begin{equation}
\Delta E(t) =  (\gamma(0)-\gamma_0) m c^2 + ec \int_0^t
\vec{F}_s\cdot\vec{\beta} dt' \label{uno0}~,
\end{equation}
where $\vec{F}_s$ is the self-force. By using vector and scalar
potential instead of the expressions for the electromagnetic
fields and then adding and subtracting the total derivative of the
scalar potential $\mathrm{d} \Phi/\mathrm{d}t$ from the integrand
in Eq. (\ref{uno0}) one gets:


\begin{equation}
\Delta E(t) =  (\gamma(0)-\gamma_0) m c^2 - e(\Phi(t)-\Phi(0)) + e
\int_0^t {\partial(\Phi - \vec\beta\cdot\vec A)\over{\partial t'}}
dt' .\label{uno}
\end{equation}
Note that the first term on the right hand side of Eq. (\ref{uno})
is simply the initial kinetic energy deviation from the design
kinetic energy, the second is due to space charge while the third
is due to longitudinal self-interactions others than space-charge.

On the other hand, using the same formalism:

\begin{equation}
F_\bot = - e {\partial(\Phi - \vec\beta\cdot\vec A)\over{\partial
r}} - e {d A_r \over {c dt}} + e {\beta_s A_s\over{r}} \label{due}
\end{equation}
The terms in Eq. (\ref{uno}) and Eq. (\ref{due}) are rearranged in
\cite{RUI} thus resulting in the following expression for Eq.
(\ref{transverse})

\begin{equation}
{d\over{dt}}\left(\gamma m {d r\over{dt}} \right) - \gamma_0 m c^2
\left({\beta_s^2\over{r}}-{\beta_s\beta_0\over{R}}\right) = G_0 +
G_\| + G_r + G_c \label{trui}
\end{equation}

where

\begin{equation}
G_0 = \beta_s^2{(\gamma(0)-\gamma_0) m c^2 +
e\Phi(0)\over{r}}\label{G1}
\end{equation}

\begin{equation}
G_\| = {e\beta_s^2 \over{r}} \int_0^t {\partial(\Phi -
\vec\beta\cdot\vec A)\over{\partial t'}} \mathrm{d}t'\label{G2}
\end{equation}

\begin{equation}
G_r = -e {\partial(\Phi - \vec\beta\cdot\vec A)\over{\partial r}}
- e {d A_r \over {c dt}} \label{G3}
\end{equation}

\begin{equation}
G_c = e\beta_s {A_s - \beta_s\Phi\over{r}}\label{G4}
\end{equation}

According to the definition in \cite{RUI,BOHN,DERUI}, the
cancellation effect consists in the recognition that $G_c \simeq
0$ and can be therefore neglected in simulations. Note that this
is obvious in the steady state case in the zeroth perturbation
order in the self-fields, when $A_s=\beta_s \Phi$ strictly. In
Fig. 2 of \cite{RUI} it is shown, by means of macroparticle
simulations, that $G_c$ is "always"\footnote{The simulation
results in \cite{RUI} show indeed the results for a beam in an arc
of a circle.} negligible. The importance of this finding, in the
view of the authors of \cite{RUI,BOHN,DERUI}, is that the last
term of Eq. (\ref{trui}), i.e. the expression in Eq. (\ref{G4}),
is negligible and free from logarithmic singularity in $x$, while
$F_\perp$ is singular (again, when a specific distribution which
allows singularities is selected to describe the particle beam). A
proof of the independence of Eq. (\ref{G4}) on the logarithm of
$x$ in spite of the dependence of the total force is given in
\cite{RUI}, in that this term does not share the same spread as a
function of the position inside the bunch with the total force
(see \cite{RUI}, Fig. 1). This fact is seen as full of deep
physical meaning and it is reported to be of great practical
interest since it would permit major simplifications when taken
into account in a computational scheme \cite{RUI,BOHN,DERUI}.
Moreover, it has been said that the cancellation effect  "opens
new possibility for theoretical investigations such as
two-dimensional analysis based on Vlasov's equation" (quoted from
\cite{BOHN}).

\section{Our arguments against the relevance of the cancellation effects \label{sec:attack}}

While we agree with the findings in \cite{LEE} concerning the
tail-head interaction within a coasting beam in a circular motion,
we disagree on the fact that such results can be extended in an
useful way in the more generic situation of a bunched beam moving
in an arc of a circle. Moreover, it must be underlined that the
findings in \cite{LEE} do not account for the head-tail part of
the interaction, arising from electrons in front of the test
particle. In the following we show that the cancellation effect is
indeed artificial and it has no important physical meaning.
Furthermore we show that, from a practical viewpoint, in building
a simulation, there is no reason to prefer the cancellation
approach to the standard approach of calculating, separately, the
Lorenz force and the kinetic energy deviation. The latter approach
is used today by $\verb"TraFiC"^4$, a macroparticle simulation
code devoted to XFEL simulation and internationally recognized
(DESY, where it has been developed, SLAC and SPRING-8).


A feature of self-interaction which has always been forgotten in
analytical calculations and interpretation of numerical results,
is constituted by the head-tail interaction, that is that part of
the self interaction coming from particles which are situated
\textit{in front} of the test particle. CSR theory (see
\cite{SAL1}) tells that in the case of longitudinal self-fields
(linked to CSR phenomena) the head-tail part is simply related to
Coulomb repulsion, and it can be regularized away when one wants
to account for pure radiative phenomena, since this part of the
interaction is non-dissipative and linked with the space-charge
forces in a straight trajectory. The situation is completely
different when one is dealing with transverse interactions. As we
showed in \cite{OURS1}, it turns out that all the non-negligible
contribution to the head-tail interaction is, in this case, due to
a term in the acceleration field and that there is no similar
regularization mechanism which can be used as in the longitudinal
case; note that (besides the fact the head-tail term is of
accelerative nature) in the transverse case we cannot distinguish
between dissipative and non-dissipative phenomena, since no
radiation is associated to the transverse forces, at the first
order in the electromagnetic fields.

The characteristics of head-tail interactions are somehow
surprising. The magnitude of the effect turns out to be of the
same order of magnitude of the tail-head interaction. For example,
in the simple case of two particles on-orbit in a circular motion,
the tail-head interaction can be expressed as (see \cite{OURS1}):

\begin{equation}
{F_\bot} \simeq {e^2 \gamma^3 \over {4 \pi \varepsilon_0 R^2}}
\Psi(\hat{\phi})~, \label{Totexp}
\end{equation}
where $R$ is the circle radius, $\hat \phi = \gamma\phi$, $\phi$
being the retarded angle while $\Psi$ is defined by

\begin{equation}
\Psi(\hat{\phi}) = {2+\hat{\phi}^4/8 \over{\hat{\phi}
(1+\hat{\phi}^2/4)^3}} ~,\label{Phi}
\end{equation}

\noindent and the retardation condition reads:

\begin{equation}
\Delta s = (1-\beta)R\phi + {R\phi^3\over{24}}~.
\label{retsteadyappr}
\end{equation}
where $\Delta s$ is the (curvilinear) distance between the two
particles. On the other hand, the only non-negligible term in the
head-tail acceleration is of pure accelerative origin and it is
given by (see \cite{OURS1}):

\begin{equation}
{F_\bot} \simeq {e^2 \over {4 \pi \varepsilon_0 R \Delta s}}~,
 \label{Fperpht}
\end{equation}

\begin{figure}[h]
\begin{center}
\includegraphics[width=120mm]{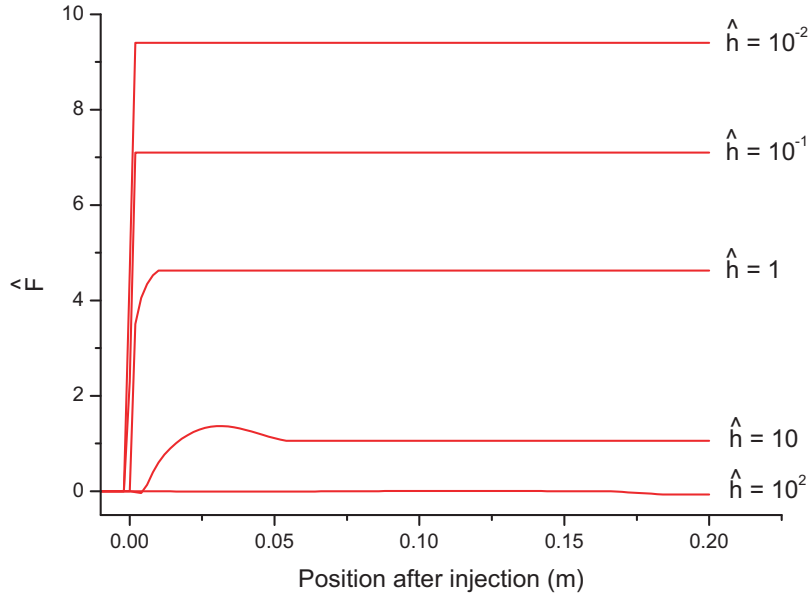}
\caption{Normalized radial force from the head of the bunch
($\Delta s < 0$) to the tail as the bunch progresses inside the
bend ($z=0$ corresponds to the injection of the test particle in
the magnet). Here $R=1~$m and $\gamma = 100$. Here we plot the
results from TraFiC$^4$ for several values of $\hat{h}=h
\gamma^2/R$, $h$ being the vertical displacement of the test
particle. The bunch length is $200 \mu m$. The test particle is
located in the middle of the bunch.\label{OutputHT}}
\end{center}
\end{figure}

\begin{figure}[h]
\begin{center}
\includegraphics[width=120mm]{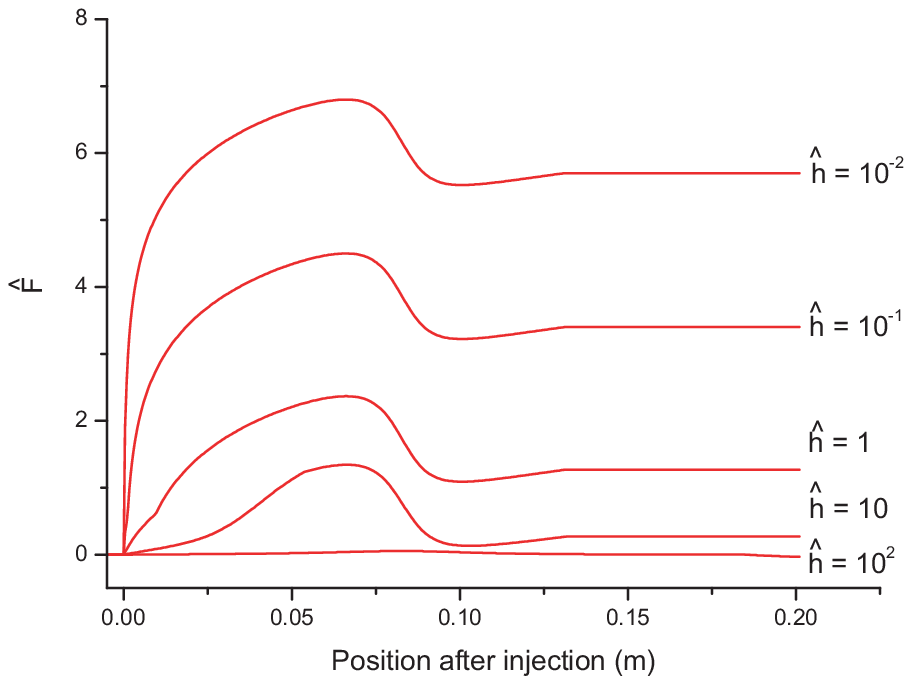}
\caption{\label{BunchTail} Normalized radial force acting on a
test particle from a bunch with rectangular density distribution
entering a hard-edge bending magnet as a function of the position
of the test particle inside the magnet. The solid lines show
analytical results; the circles describe the outcome from
TraFiC$^4$. We chose $\gamma = 100$, $R=1$ m; graphs are plotted
for several values of the parameter $\hat{h}=h \gamma^2/R$, $h$
being the vertical displacement of the test particle. The bunch
length is $200 \mu~$m. The test particle is located in the middle
of the bunch. \label{BunchTH}}
\end{center}
\end{figure}

It is easy to see, by comparison of Eq. (\ref{Totexp}) and Eq.
(\ref{Fperpht}), that the tail-head and the head-tail
contributions are of the same order of magnitude. In practical
situations, the head-tail force has been proven to be about two
times larger than the tail-head contribution: this can be seen by
direct inspection of Fig.~\ref{OutputHT} and Fig.~\ref{BunchTH},
taken from \cite{OURS2}, which show, respectively the tail-head
and the head-tail interaction exerted by a line bunch entering a
bending magnet on a particle with vertical \footnote{i.e.
orthogonal to the bending plane} displacement $h$. In the figures
a normalized parameter $\hat{h}=h\gamma^2/R$ is used, where $R$ is
the circle radius. Note that the plots show the normalized
transverse force in the radial direction, $\hat{F} = F_\bot/[e^2
\lambda/(4\pi\varepsilon_0 R)]$, $\lambda$ being the bunch density
distribution.


Let us spend some words to describe the head-tail interaction (an
exhaustive treatment can be found in \cite{OURS1,OURS2}). On the
one hand it is evident that, when $\Delta s < 0$,  the source
particle is ahead of the test electron at any time; on the other
hand it is not true that the retarded position of the source
particle is, in general, ahead of the present position of the test
particle. If $\Delta s < 0$ and, approximatively, $|\Delta s| < h$
the test particle overtakes the retarded position of the source
before the electromagnetic signal reaches it. In this case,
although we may still talk about head-tail interaction, since
$\Delta s < 0$, its real character is very much similar to the
case $\Delta s > 0$, in which the electromagnetic signal has to
catch up with the test particle. In the case of head-tail
interactions with $|\Delta s|>h$, the retardation condition has
unique features since the electromagnetic signal runs against the
test particle, and not viceversa, as in the tail-head case (or in
the head-tail case with $|\Delta s|<h$, when the electromagnetic
is running after it. The fact that the electromagnetic signal and
the test particle approach on a head-on collision is the reason
why the character of this interaction is local, in the sense that
the distance which the test particle travels between the emission
of the electromagnetic signal by a source and its reception by the
test particle is about half of the distance between the test
particle and the source, while the usual tail-head interaction has
a formation length about $\gamma^2$ times longer. This can be seen
directly comparing Fig. \ref{OutputHT} and Fig. \ref{BunchTH}.

\begin{figure}
\begin{center}
\includegraphics*[width=120mm]{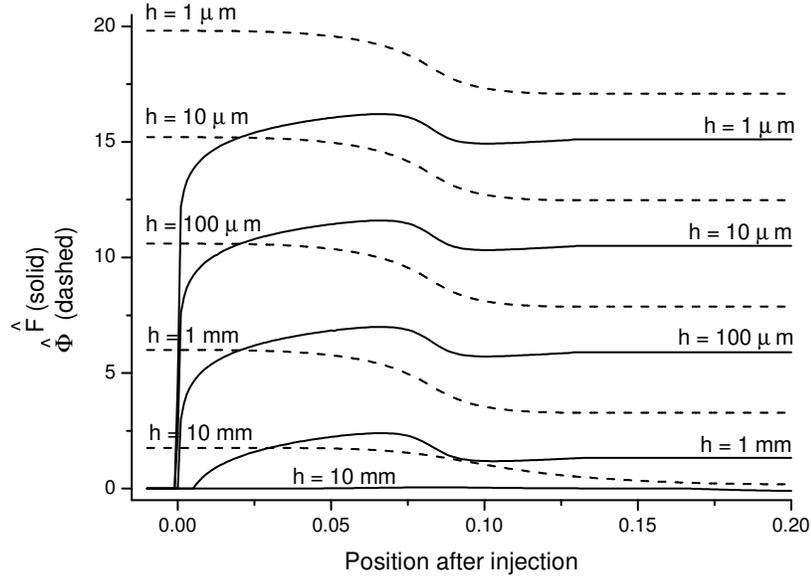}
\caption{Normalized radial force (solid line) and normalized
potential (dashed line) in the center of a 200 $\mu$m bunch as a
function of the injection position. The parameter on the curves
refers to the vertical displacement $h$ of the test particle.
\label{pot} }
\end{center}
\end{figure}

It is somehow surprising that, although simulations automatically
take care of this term, this has been overlooked for a long time
in theoretical analysis or in the interpretation of simulation
results, starting from \cite{TALM}, continuing with \cite{LEE},
\cite{DERB2} and finally \cite{RUI}.  This important term is
responsible for sudden jumps of $\hat{F}$ at the magnet entrance.
Consider for example the plot in Fig. \ref{pot}, obtained by using
the electromagnetic solver of $\verb"TraFiC"^4$. The figure shows
(solid line) the normalized radial force $\hat{F}$ felt by a test
particle (this is just the sum of Fig. \ref{OutputHT} and Fig.
\ref{BunchTH}) in the middle of the bunch and the normalized
potential $\hat{\Phi} = \Phi/[e^2 \lambda/(4\pi\varepsilon_0)]$
(dashed line) at the test particle position. The sudden jumps at
the magnet entrance are due to head-tail interaction with $|\Delta
s|>h$. Yet, before \cite{OURS1,OURS2}, there was no theoretical
explanation for this behavior. Although Fig. \ref{pot} refers to
displacement in the vertical plane, it is obvious that a similar
displacement would be found in the horizontal one too, signifying
a logarithmic dependence on $x$.

%

Now that the head-tail interaction has been introduced, it is
possible to give a very simple argument against the meaningfulness
of the cancellation scheme. In \cite{RUI} attention is drawn only
on $G_r$ and $G_c$, while no word is spent about $G_0$ and $G_\|$.
As already said, in Fig \ref{pot} we also plotted (dashed lines)
the normalized potential $\hat{\Phi}$ in the center of the bunch
for different vertical displacements $h$ as calculated by
$\verb"TraFiC"^4$. As it is seen, the potential does not share the
same formation length of the head-tail force, which clearly
demonstrates that no effective cancellation can take place between
$F_\bot$ and $\Delta E$. This fact can be shown in the same
terminology used in \cite{RUI}. Even if $G_c$ is negligible and
$G_r$ is centripetal, there is a huge, dominant centrifugal
contribution on the right hand side of Eq. (\ref{trui}): this is
$G_0$ which includes the term $\Phi(0)/R$. Since the head-tail
interaction has local nature, we can say that, due to the
rearrangement proposed in \cite{RUI}, $G_0$ now includes all the
head-tail interaction part (the sharp time dependence of the
head-tail interaction being masked in the other terms of Eq.
(\ref{due}), i.e. in the terms $G_\|$, $G_r$ (and $G_c$)).

Actually our result shows that $G_c \sim 0$, alone, is
uninteresting, that is, the cancellation effect as defined in
\cite{RUI} is uninteresting. There will be anyway a strong
head-tail contribution which is accounted in the first term $G_0$.
Since this term depends on the position of the test particle
inside the bunch it will lead to emittance growth both in the
transient and in the steady state case. This term has not been
taken into account in \cite{TALM} nor in \cite{LEE}; it has been
automatically included in the simulations in \cite{RUI}, because
the computational scheme is correct, although unnecessarily
complicated, but then it has been disregarded in the analysis of
the results. Note that the first term $G_0$ would give a spread
similar to that of $F_\bot$ in Fig. 1 of reference \cite{RUI} if
plotted for a bunch with a given energy spread, since it depends
logarithmically on $x$ (at least in the 1D model with the test
particle displaced horizontally). In other words, if there is a
field singularity, it is there also in the right hand side of the
equation of motion and it cannot be cancelled away. Note that
there is no possibility of cancelling $G_0$ by choosing a certain
initial condition for the bunch. In fact the freedom of choice of
the initial condition refers to the kinetic energy deviation from
the design energy, and not to the deviation of the total energy
(kinetic energy summed to the potential energy) from the design
kinetic energy.

This shows that there is no theoretical nor practical reason for
adopting the cancellation scheme as a preferred one with respect
to the usual scheme of calculating separately $F_\bot$ and $\Delta
E$. In fact, from a practical viewpoint one will encounter the
same computational difficulties and from a theoretical viewpoint
two easily understandable terms are mixed up into four terms (of
which three survive and one is approximately cancelled) with a
more involved physical interpretation. Not to mention that the
cancellation method can be extremely confusing. The latter is a
statement, not an opinion; in the LCLS Design Study Report
\cite{l} we read that all contributions to the the transverse
emittance growth are due to the centripetal force "which
originates from radiation of trailing particles and depends on the
local charge density along the bunch. The maximum force takes
place at the center of the bunch and its effect on the transverse
emittance is estimated in the reference."; the reference in the
quoted passage is the work by Derbenev and Shiltsev \cite{DERB2}.
Here the head-tail contributions are completely neglected, because
of the confusion coming form the cancellation issue: in fact the
(centripetal) force in the quote is the third term in Li's
treatment ($G_r$, the only one considered) while the first term
$G_0$, containing the (centrifugal) head-tail interaction and the
second, $G\|$, are completely neglected. This is an example where
on the one hand the code $\verb"TraFiC"^4$ was correctly used
giving correct results, but where, on the other hand, its results
were completely misunderstood. Nevertheless, as we already said,
the head-tail interaction is a few times larger than the tail-head
interaction, and, since it depends on the position of the test
particle along the bunch, it is obviously responsible for
normalized emittance growth.

The cancellation scheme as in \cite{RUI,BOHN,DERUI}, actually
reduces to the following: get a quantity $A+B$ and express
$A=A_1+A_2$, $B=B_1+B_2$ where $A_2\simeq -B_2$, leaving large
part of $A$ (and even singular quantities) in $A_1$, complicate
the situation furthermore breaking $A_1+B_1$ in several parts and
then claim that the cancellation $A_2+B_2\simeq 0$ is an important
finding. This can be done with any pair of quantities $A$ and $B$,
and it is, in our view, completely trivial and uninteresting. As a
final remark, one could have foreseen that the cancellation effect
cannot be of any use since the transverse head-tail force, which
has local character cannot be effectively cancelled by the energy
deviation, which depends upon all the trajectory.

\section{Conclusions}

In this paper we proved that the cancellation effect is an
artificial one. In the right hand side of the transverse equation
of motion the Lorentz force and the kinetic energy deviation from
the design energy, which are clearly physically meaningful
quantities, are combined together to give four different terms of
difficult physical interpretation. The cancellation effect deals
only with one of these terms, still leaving other three important
contributions to be evaluated. We found in \cite{OURS1,OURS2} that
the head-tail interaction is, in practical situations, a few times
larger than the tail-head interaction. This fact, automatically
included in computer simulation, has always been forgotten in
analytical considerations, starting from \cite{TALM} on. The
formation length of the head-tail interaction with $|\Delta s|>h$
is approximately half of the bunch length, which means that, for
practical purposes, the head-tail interaction has a local nature.
This explains the sudden "jump" in the total Lorentz force seen as
a bunch enters a bending magnet. On the other hand, the kinetic
energy deviation from the design energy has much a larger
formation length. This fact alone is sufficient to prove that the
cancellation effect is an artificial one, in that an important
contribution (the head-tail interaction) cannot be compensated by
the potential term in the kinetic energy deviation.

When we state that the cancellation effect is artificial, we do
not mean that it is, strictly speaking, wrong; we simply state
that it is not useful at all and that there is no theoretical nor
practical reason to prefer this method to the more straightforward
approach of calculating, separately, the transverse Lorentz force
and the kinetic energy deviation term (as, for example, by means
of the code $\verb"TraFiC"^4$). In fact, from a practical
viewpoint one will encounter the same computational difficulties
and from a theoretical viewpoint two easily understandable terms
are mixed up into four terms (of which three survive and one is
approximately cancelled) with a more involved physical
interpretation.

\section{Acknowledgements}

We wish to thank Martin Dohlus for providing numerical
calculations using the code $\verb"TraFiC"^4$. Also, thanks to
Joerg Rossbach and Marnix van der Wiel for their interest in this
work.

\end{document}